\documentclass[prb,twocolumn,amsmath,amssymb,aps
]{revtex4-2}
\usepackage{ulem}
\usepackage{lipsum}
\usepackage{graphicx}
\usepackage{dcolumn}
\usepackage{bm}
\usepackage[colorlinks=true]{hyperref}
\makeatletter
\newcommand*{\rom}[1]{\expandafter\@slowromancap\romannumeral #1@}

\usepackage {xcolor}

\def\bef{\begin{framed}}
	\def\eef{\end{framed}}
\def\be{\begin{equation}}
	\def\ee{\end{equation}}
\def\ber{\begin{eqnarray}}
	\def\eer{\end{eqnarray}}

\def\nn{\nonumber}

\begin{document}
	
\title{Control of spin-wave polarity and velocity using a ferrimagnetic domain wall}
\date{\today}

\author{Ehsan Faridi$^1$} 
\author{Giovanni Vignale$^{1,2}$} 
\author{Se Kwon Kim$^3$}
\email{sekwonkim@kaist.ac.kr}

\affiliation{$^1$Department of Physics and Astronomy, University of Missouri, Columbia, Missouri 65211, USA\\$^2$ The Institute for Functional Intelligent Materials (I-FIM), National University of Singapore, 4 Science Drive 2, Singapore 117544\\$^3$Department of Physics, Korea Advanced Institute of Science and Technology, Daejeon 34141, Korea}

\begin{abstract}
We present a theoretical study of the scattering of spin waves by a domain wall (DW) in a ferrimagnetic (FiM) spin chain in which two sublattices carry spins of unequal magnitudes. We find that a narrow, but atomically smooth FiM DW exhibits a different behavior in comparison with similarly smooth ferromagnetic and antiferromagnetic DWs due to the inequivalence of the two sublattices. Specifically, for sufficiently weak anisotropy, the smaller spin at the center of the DW is found to become precisely normal to the easy-axis, selecting an arbitrary direction in the $xy$-plane and thereby breaking the U(1) spin-rotational symmetry spontaneously. This particular form of a FiM DW does not occur in antiferromagnetic systems and is shown to lead to a strong dependence of spin wave scattering pattern on the state of polarization of the spin wave, which can be either right-handed or left-handed, suggesting the utilization of such a narrow DW as a spin-wave filter. Moreover, we find that in the case of an atomically sharp DW, where all the spins point either up or down due to strong easy-axis anisotropy and therefore the polarization of the spin wave is conserved upon transmission, the wave vector of the spin wave changes after passing through the DW leading to a change in the group velocity of the spin wave. This change of the wave vector indicates the acceleration or deceleration of the spin waves and thus a sharp FiM DW could serve as a spin wave accelerator or decelerator in spintronics devices, offering a functionality absent in a ferromagnetic and an antiferromagnetic counterpart. Our results indicate that FiM spin textures can interact with spin waves distinctly from ferromagnetic and antiferromagnetic counterparts, suggesting that they may offer spin-wave functionalities that are absent in more conventional magnets.
 \end{abstract}
 
 \maketitle
	
	\section{\label{sec:level1}Introduction}
Ferrimagnets (FiMs), which consist of two unbalanced magnetic sublattices that are aligned antiferromagnetically, have emerged in spintronics as candidates that can offer advantages of both ferromagnets
and antiferromagnets~\cite{kim2022ferrimagnetic,zhang2023ferrimagnets,dionne1975review,kim2020fast,ohnuma2013spin}. The inequivalent magnetic sublattices leads to a small net magnetization which can be detected and manipulated by previously established techniques used for ferromagnets~\cite{hennecke2019angular,geprags2016origin,siddiqui2018current,ding2020identifying,polishchuk2018spin,maier2017tunable,shiomi2014interface,huebl2013high,liensberger2019exchange,ganzhorn2016spin}. Furthermore, similar to antiferromagnets, FiMs support two types of spin waves, right-handed (RH) and left-handed (LH) ones, offering a useful degree of freedom for spin wave manipulation, whereas ferromagnets can host only an RH mode~\cite{ivanov2019ultrafast,okamoto2020flipping,wu2024magnon}. These simple but compelling features have recently triggered a surge of interest in FiMs, including the studies of energy-efficient and
     ultrafast optical manipulation of FiMs~\cite{serga2010yig}, the dynamics of a domain wall (DW) in a FiM by spin transfer torques~\cite{kim2020dynamics,donges2020unveiling,jing2022field,lisenkov2019subterahertz,martinez2019current}, and the application of a FiM DW as a wave-guide in spintronics devices~\cite{liu2022handedness}.  

     Although spin waves and DWs in FiMs have been studied in the past~\cite{jin2021domain,kim2020distinct,oh2019bidirectional,oh2017coherent,kim2017fast,funada2019spin,haltz2019strong,lan2017antiferromagnetic,cheng2016antiferromagnetic,zhang2020gate}, one aspect of FiMs that has not been addressed so far is the interaction of a spin wave with a FiM DW in a regime where the width of the DW is as narrow as a few lattice constants, which can be realized in FiMs with magnetic anisotropy comparable to the exchange energy {\color{green}\cite{barbara1994magnetization}}. An analogous problem for antiferromagnets, i.e., the interaction between spin waves and a narrow antiferromagnetic DW has been studied by us~\cite{faridi2022atomic}, where it was shown that an atomically sharp DW can interact with spin waves differently from a smooth DW~\cite{kim2014propulsion,tveten2014antiferromagnetic}. More specifically, we found that when the ratio of anisotropy to the exchange constant is less than a certain critical value, both RH and LH spin waves are transmitted through the DW with the same probability. However, when the anisotropy is greater than the critical value, the DW becomes atomically sharp, and LH and RH spin waves are transmitted with different amplitudes,
     depending on the orientation of the spins at the domain boundaries.   
     
     In this work, we address the problem of the interaction of spin waves and narrow DWs in FiMs and identify differences in the FiM case from the ferromagnetic and antiferromagnetic cases. More specifically, we show that a smooth FiM DW in the case of the small anisotropy exhibits a feature that is absent in the antiferromagnetic case: the transmission coefficients for RH and LH spin waves are different even for a smooth DW. In addition, we find that in the case of an abruptly sharp FiM DW, the wave vector of the spin wave changes after passing through the DW, leading to a change in the group velocity of the spin wave, indicating acceleration or deceleration of the latter.  This is a consequence of the fact that RH and LH waves have different dispersions in the two sides of the DW, which is distinct from the antiferromagnetic case in which they have identical dispersions in the two sides of the DW in the absence of an external magnetic field.
     
     The rest of the paper is organized as follows. In Sec.~\rom{2}, we describe our model and the theoretical formulation of the problem. In Sec.~\rom{3} , we present our results for the transmission and reflection coefficients of spin waves with a FiM DW and demonstrate the spin-wave acceleration and deceleration effect. In Sec.~\rom{4}, we conclude our work by summarizing the main results and offering a future outlook.

\section{Theoretical Formulation}	 

In this section, we present the theoretical formulation that allows us to investigate the interaction of spin waves on top of a DW.

	\subsection{Structure of a FiM DW}
 
We consider a FiM spin chain with easy-axis anisotropy, where the spins of two sublattices are antiferromagnetically coupled and have different magnitudes. The pertinent Hamiltonian is given by
		\begin{equation}
			\mathcal{H}=J\sum_{n}\textbf{S}_n \textbf{.} \textbf{S}_{n+1}-\textit{D}\sum_{n}({\textbf{S}^{ z}_n})^2 \, ,
		\end{equation}	
where $\textbf{S}_n$ is the spin angular momentum at site $n$ and its magnitude alternates between $S_A$ and $S_B$. The first term in the Hamiltonian represents the exchange interaction between two consecutive spins with $J>0$,
and the second term is the easy-axis anisotropy with $D>0$. 
 In this Hamiltonian, we neglect nonlocal dipolar interaction, which can be justified for FiMs when their net magnetizations are sufficiently small~\cite{kim2021current,finley2020spintronics,haltz2022quantitative,mishra2017anomalous}.

To obtain a static DW solution, we minimize the energy with respect to the polar angle $\theta_n$ of $\textbf{S}_n = (\sin \theta_n \cos \phi_n, \sin \theta_n \sin \phi_n, \cos \theta_n)$ while setting $\phi_n = 0$ without loss of generality, which results in the following equation:
\begin{equation}
\sin(\theta_n-\theta_{n+1})+\sin(\theta_n-\theta_{n-1})-\dfrac{D}{J}\frac{S_n^2}{S_{A}S_{B}}\sin2\theta_n =0 \, .
\label{eq:dw-eq}
\end{equation} 
Here, we assume that $S_n$ takes $S_A(S_B)$ for even (odd) site,  $n$. The solution of the equation depends on the ratio of anisotropy to exchange constant $D/J$, as well as the ratio of the magnitudes of spins on the two sublattices i.e., $S_A/S_B$. For the special case of $S_A = S_B$, i.e., the antiferromagnetic case, it is known that there are two types of DWs depending on $D/J$: Eq.~(2) yields an antiferromagnetic Walker-type smooth DW for small anisotropy $D/J\ll 1$ and an atomically sharp DW for large anisotropy $D/J \geq2/3$~\cite{barbara1994magnetization,faridi2022atomic}.  

\begin{figure}
	\centering
\includegraphics[width=85mm]{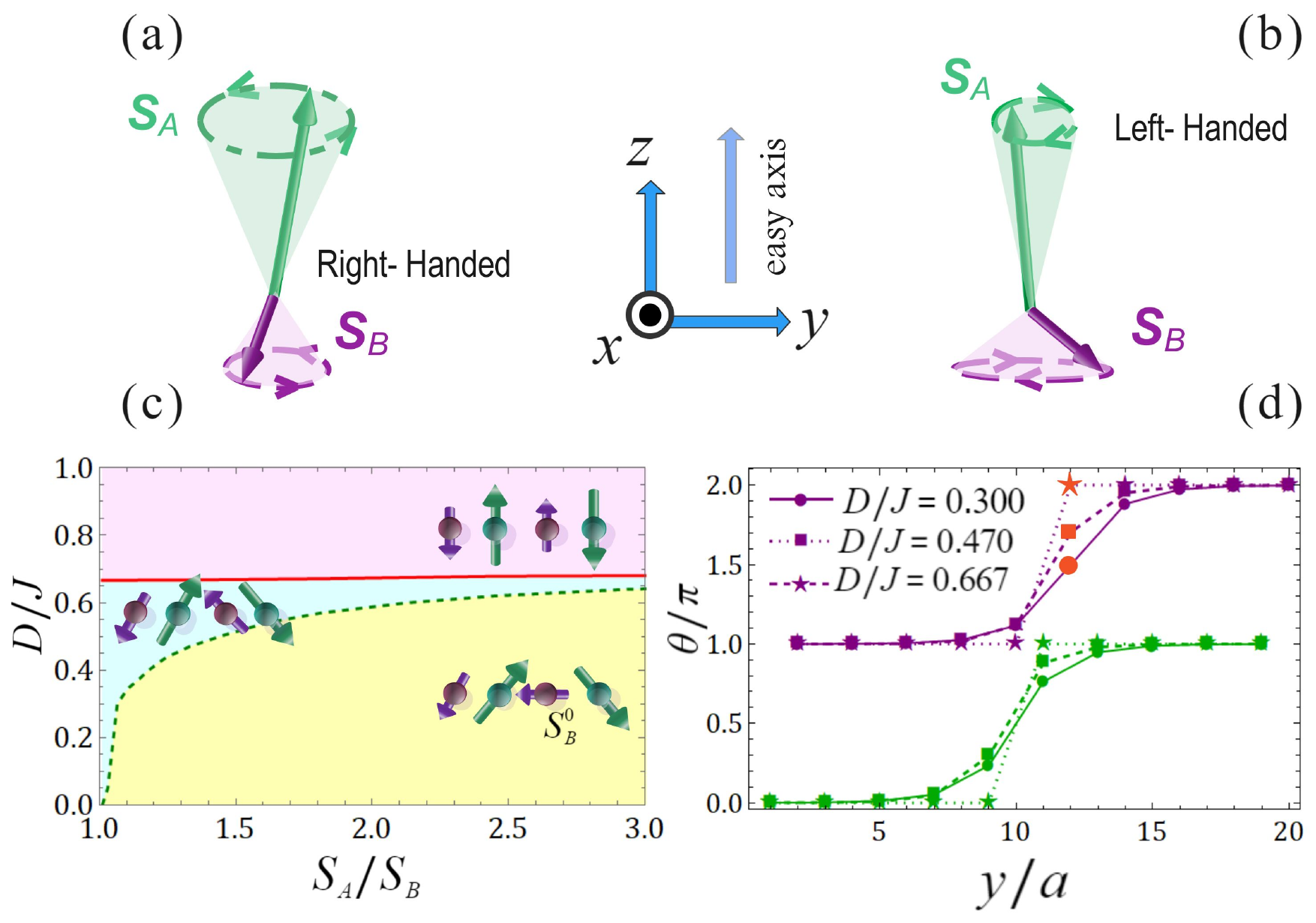}
\caption{Schematic illustration of the motion of the two sublattice spins $\mathbf{S}_A$ and $\mathbf{S}_B$ for (a) a right-handed spin wave and (b) a left-handed spin wave.  (c) Phase diagram of the FiM DW spin configuration in the parameter space spanned by the relative spin magnitude $S_A / S_B$ and the relative strength of the easy-axis anisotropy compared to the exchange energy $D/J$. For each phase, the four consecutive spins located at the center of the spin chain that contains a FiM DW are depicted. The spin at the center of the DW is denoted by $\mathbf{S}_B^0$. In the pink region where the easy-axis anisotropy is very strong, all the spins are collinear with the easy-axis and therefore the DW is atomically sharp. In the green region where the easy-axis anisotropy is intermediate, the spins are neither collinear nor perpendicular to the easy-axis. In the yellow region where the easy-axis anisotropy is  weak, the central spin $\mathbf{S}_B^0$ is perpendicular to the easy-axis. The boundary between the pink and the green region is given in Eq.~(\ref{eq:dc}) and the boundary between the green and the yellow regions is obtained numerically by solving Eq.~(\ref{eq:dw-eq}). (d) The spin configuration of a DW for $S_A=5/2$ and $S_B=2$ for three different values of $D/J$. The green and the purple points represent the polar angle $\theta$ with the $z$-axis for the sublattices A and B, respectively. The points colored orange represent the polar angle of the central spin $\mathbf{S}_B^0$. Note that for $D/J = 0.3$, the polar angle of the central spin $\mathbf{S}_B^0$ is $3 \pi/2$, meaning that the central spin is perpendicular to the easy-axis.}
	\label{DW}
\end{figure}

When the spins of the two sublattices are different, it turns out that there are three different types of DWs, differing from the antiferromagnetic case exhibiting only two types of DWs. Without loss of generality, we assume $S_A > S_B$ throughout. The type of a DW can be characterized by the angle formed by the central spin on the sublattice B (denoted by $\mathbf{S}_B^0$) with the $z$-axis as shown in Fig.~\ref{DW}(c). In the yellow region (small $D/J$), the direction of $\mathbf{S}_B^0$ is normal to the easy-axis, i.e., lies in the $xy$ plane. As $D/J$ increases the DW becomes narrower, but the overall shape of the DW remains unchanged. At the first critical value of $D/J$, the direction of $\mathbf{S}_B^0$ begins to deviate from the normal to the easy-axis due to the effect of the stronger easy-axis anisotropy, entering the phase corresponding to the green region of Fig.~\ref{DW}(c). Finally, for $D/J$ larger than the second critical value, the DW becomes abruptly sharp, entering the phase corresponding to the pink region of the figure: all the spins, including the center spin $\mathbf{S}_B^0$, are either parallel or antiparallel to the $z$-axis. In contrast to ferromagnetic or antiferromagnetic domain walls, which become abruptly sharp at a specific value of $D/J = 2/3$ {\color{green}\cite{barbara1994magnetization}}, the critical value of $D/J$ above which a FiM DW becomes atomically sharp, denoted by $d_c$, depends on the ratio $S_A/S_B$,  but only very weakly, increasing from $d_c=2/3$ for $S_A/S_B=1$ to  $d_c \approx 0.679$ for $S_A/S_B=3$. The critical value of $d_c$ can be obtained analytically by identifying the condition under which the zero-frequency Goldstone mode disappears. An analytic formula for $d_c$ is derived in Appendix A.  

Note that the phase colored yellow with $\mathbf{S}_B^0$ exactly normal to the easy axis is not present for the antiferromagnetic case $S_A / S_B = 1$, which agrees with the known result for the antiferromagnetic case~\cite{faridi2022atomic}.  The phase boundary between the green and the pink region is obtained numerically by solving Eq.~(\ref{eq:dw-eq}). Demonstrating the existence of three distinct types of DWs in FiMs, as shown in Fig.~\ref{DW},  is one of our main findings in this paper.

\bigskip
	
\subsection{Equation of motion}	

Starting from the equilibrium configuration of a DW, we construct the equation of motion, $d\textbf{S}_n/dt=-\textbf{S}_{n}\times (-\delta \mathcal{H}/\delta \textbf{S}_{n})$, governing the time evolution of an individual spin around its equilibrium orientation $\textbf{S}_{n,0}=S_{n}(0,\cos\theta_n,\sin\theta_n)$ at site $n$. To represent small excitations around the equilibrium configuration, we transform from a global reference frame $(x,y,z)$ (the ``lab" frame) into a local reference frame $(X,Y,Z)$, 
	which is rotated
	about the $x$-axis in such a way that the local $Z$-axis coincides
	with the local orientation of $S_n$ at equilibrium, as has been done for antiferromagnets~\cite{faridi2022atomic}.
	The relation between the components of the spin in the
	local coordinate system  and in the global coordinate
	system is
		\begin{equation}
		\begin{pmatrix}
			S_{ nx}  \\S_{ny} \\S_{nz}
		\end{pmatrix}=	
		\begin{pmatrix}
			1&0&0 \\
			0&\cos{\theta_n}& \sin{\theta_n}\\ 0 & -\sin{\theta_n} & \cos{\theta_n} 
		\end{pmatrix}
		\begin{pmatrix}
			S_{ nX} \\S_{nY} \\S_{ nZ}
		\end{pmatrix} \, .
	\end{equation}
By keeping linear terms in the oscillation amplitudes of spins, $S_{nX}$ and $S_{nY}$ and by considering monochromatic solutions at the frequency $\omega$, the equation of motion takes the following form:

\begin{equation} 
	\hbar\omega S_{n i} = \sum_{n' j}H_{n i,n' j} S_{n' j}\,.
\label{schero_spin_wave}
\end{equation}
Here, the diagonal part of the spin wave Hamiltonian is expressed as:
\begin{eqnarray}
	H_{n i,n j}&=& -J(S_{n-1}c_{n-1}  +S_{n+1}c_n)[\sigma_z]_{i j} \nonumber \\  &+&   DS_n(2\cos^2 \theta_n-\sin^2\theta_n) [\sigma_z]_{i j}\nonumber\\
	&+& DS_n\sin^2\theta_n[i\sigma_y]_{i j}\,,
\end{eqnarray}
and the off-diagonal part is given by
\begin{eqnarray}
	H_{n i,(n+1) j }&=&JS_n\left\{\frac{1+c_n}{2} [\sigma_z]_{i  j}+\frac{1-c_n}{2}[i\sigma_y]_{i j}\right\}\,,\nonumber\\
	H_{n i,(n-1) j}&=&JS_n\left\{\frac{1+c_{n-1}}{2} [\sigma_z]_{i j}+\frac{1-c_{n-1}}{2}[i\sigma_y]_{i j}\right\}.\nn\\
\end{eqnarray}
where $i$ and $j$ take the values in $\{-,+\}$,  $S_{n\pm}=S_{nX}\pm i S_{nY}$   are the chiral components of the spin deviation, $c_n=\cos(\theta_n-\theta_{n+1})$, and $\sigma_i$'s are the Pauli matrices.

 \begin{figure*}
 	\centering
 	\includegraphics[width=2\columnwidth]{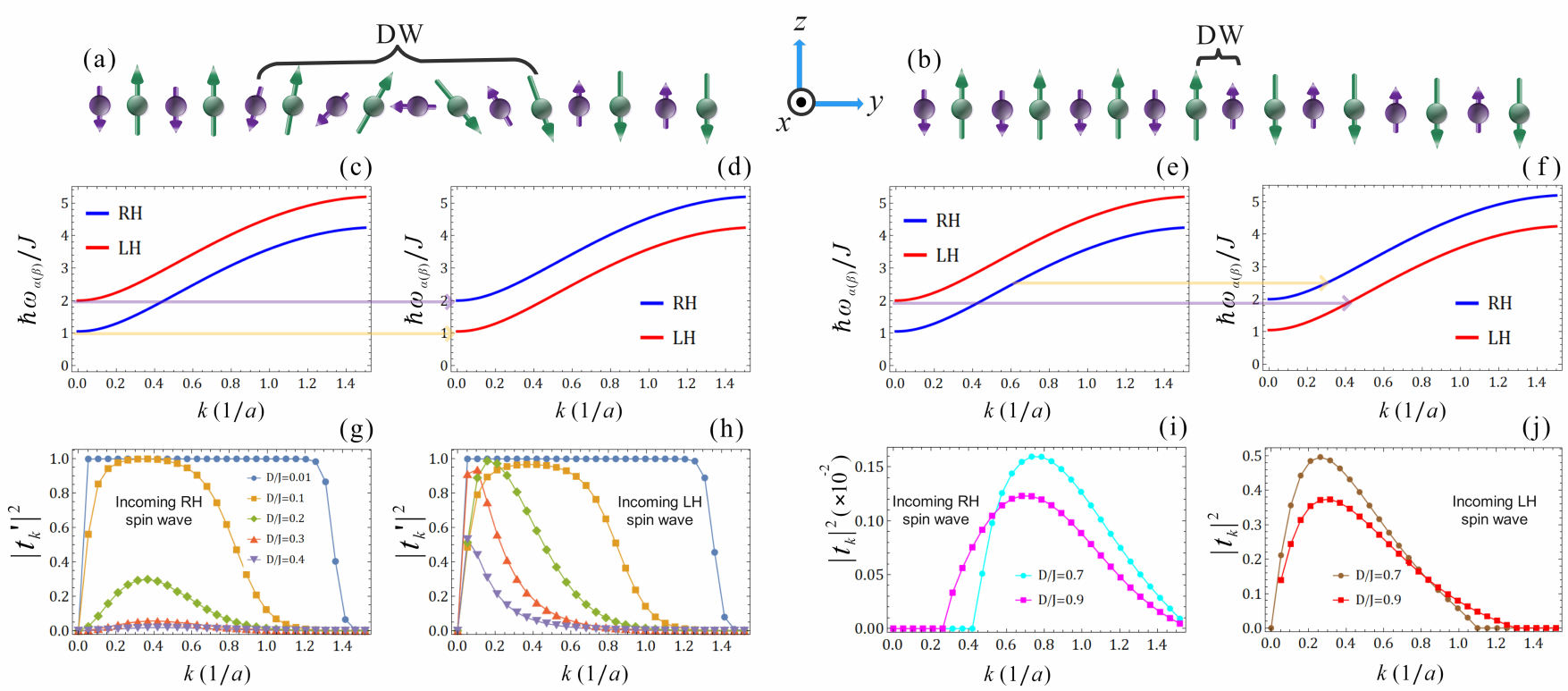}
	\caption{ 
  (a) Schematics of a  narrow FiM DW where the spin at the center of the DW is perpendicular to the easy axis. (b) Schematics of an abruptly sharp FiM DW where all the spins are either parallel or antiparallel to the easy axis. (c) Dispersion relations [Eqs.~(\ref{eq:omega-alpha}) and (\ref{eq:omega-beta})] of an RH (blue curve) and an LH (red curve) spin waves on the left side of the DW. (d)  Dispersion relation of an RH (blue curve) and an LH (red curve) spin waves on the right side of the DW where the order parameter is reversed relative to the left side. Note that the polarization of the spin wave changes, but the wave vector of the spin wave does not change after the transmission (connected by purple and yellow lines from (c) to (d)). (e), (f) The dispersion relations of spin waves on the left and the right of the DW in the case of an abruptly sharp DW. Note that the polarization of the spin wave does not change, but the wave vector changes after the transmission in order to keep the frequency constant, as illustrated by the purple and yellow lines. (g),(h) Transmission coefficients [Eq.~(\ref{eq:right})] of an RH and an LH spin wave incident from the left upon a FiM DW for different values of $D/J$, respectively. (i), (j) Transmission coefficients of an RH and an LH  spin wave incident from the left upon an abruptly sharp FiM DW, respectively.}
 	\label{sharp}
 \end{figure*}

\subsection{Spin waves}

Before considering the scattering of spin waves with a DW, we first discuss the eigenvalues and eigenvectors of spin waves in a ground state of a FiM spin chain characterized by $\theta_n=0$ for even sites (sublattice A) and $\theta_n=\pi$ for the odd sites (sublattice B). There are two solutions to Eq.~(\ref{schero_spin_wave}), which are orthogonal eigenvectors denoted by $\alpha$- and $\beta$-mode. The $\alpha$-mode solution is given by
 \be  u_{k,A}^{\alpha}=\begin{pmatrix} 
 	1\\0  
 \end{pmatrix},
 ~ u_{k,B}^{\alpha}=  \dfrac{-2JS_{B}\cos (ka)}{(S_{A}+S_{B})(D+J)+\hbar\omega_0(k)}\begin{pmatrix} 
 	0\\1  
 \end{pmatrix},
 \ee
with the eigenvalue
 \be
 \hbar\mathcal{\omega}_{\alpha}(k)=(D-J)(S_{A}-S_{B})+\hbar\omega_0(k) \, , 
 \label{eq:omega-alpha}
 \ee
 and
  \be
 \hbar\omega_0(k)=\sqrt{(S_{A}+S_{B})^2(J+D)^2-4S_A S_{B}J^2\cos^2(ka)} \, ,
 \ee 
where the first and the second components are the oscillation amplitudes of spins at site A and B, respectively. For the $\beta$-mode, we have
 \be 
  u_{k,A}^{\beta}=\begin{pmatrix} 
 	0\\1  
 \end{pmatrix},
 u_{k,B}^{\beta}=  \dfrac{-2JS_{B}\cos (ka)}{(S_{A}+S_{B})(D+J)-\hbar\omega_0(k)}\begin{pmatrix} 
 	1\\0  
 \end{pmatrix},
 \ee
with the eigenvalue
 \be
 \hbar\mathcal{\omega}_{\beta}(k)=-(D-J)(S_{A}-S_{B})+\hbar\omega_0(k) \, .
\label{eq:omega-beta}
 \ee
For $S_A=S_B$ the two eigenmodes become degenerate and we recover antiferromagnetic spin-wave spectrum $\omega_\alpha(k) = \omega_\beta(k) = \omega_0(k)$. A schematic illustration of two FiM eigenmodes are shown in Fig.~ \ref{DW}.

Now let us consider a DW and spin waves on top of it. Before delving into the formal analysis, we discuss qualitative features of spin waves on the two sides of the DW, each of which can be considered as homogeneous FiM domains. See Fig.~\ref{sharp}(a) and Fig.~\ref{sharp}(b) for the illustration of the narrow, but atomically-smooth DW and the atomically sharp DW, respectively. Note that on the left side of the DW, where the spins at the sublattice A(B) are oriented in the $+z(-z)$ direction the $\alpha$-mode has an RH polarization and the $\beta$-mode has an LH polarization in the global reference frame. On the right side of the DW, the orientation of spins is reversed, thus the $\alpha$-mode represents an LH  polarization in the global reference frame, while the $\beta$-mode represents a RH polarization.

We remark here that the frequency split between RH and LH spin waves occurs also in antiferromagnets in the presence of a uniform magnetic field collinear to the order
parameter \cite{shen2020driving} similar to the FiM case, but there is one important difference between the FiM and the antiferromagnetic cases. In the case of the antiferromagnetic system, the magnetic field is the same on the left and on the right side of the DW and therefore the splitting has the same sign on the two sides, meaning that, for example, RH spin wave has a lower frequency and LH spin wave has a higher frequency on both sides of the DW. In the case of the FiM, splitting of the frequency between the RH and the LH spin waves arises, not from the external magnetic field, but from the net magnetization intrinsic to the FiM, which has opposite signs on the two sides of the DW. Therefore the RH (LH) spin wave has a lower (higher) frequency on the left but a higher (lower) frequency on the right side of the DW.

Now, let us turn to the formulation of the scattering of spin waves incident on the DW from the left. The solutions to the scattering problem have the form of a plane wave. On the left side of the DW, the spin-wave solution is a superposition of an incoming spin wave with a certain polarization, say $\alpha$, and a reflected spin wave of the same polarization with amplitude $r$ and a reflected spin wave of opposite polarization with amplitude $r'$:
 \ber
\psi_k(n) &=& e^{ikn}\left\{u_{k,A}^{\alpha}\delta_{\bar n,0} +u_{k,B}^{\alpha}\delta_{\bar n,1}\right\}\nn\\
&+&re^{-ikn}\left\{u_{k,A}^{\alpha}\delta_{\bar n,0}+u_{k,B}^{\alpha}\delta_{\bar n,1}\right\}\nn\\
&+&r'e^{-i\kappa n}\left\{u_{\kappa,A}^{\beta}\delta_{\bar n,0}+u_{\kappa,B}^{\beta}\delta_{\bar n,1}\right\} \, ,
\label{eq:left}
\eer
for $n \leq 0$. Here, $\kappa$ is the wave vector of the reflected spin wave with a reversed polarization but with the same frequency as the incident wave. On the right side of the DW, $n > 2N$, where $2N$ is the number of spins
 inside the DW, the transmitted spin wave has the form 
 \ber
 \psi_k(n) &=&  
 t'e^{ikn}\left\{u_{k,A}^{\alpha}\delta_{\bar n,0}+u_{k,B}^{\alpha}\delta_{\bar n,1}\right\}\nn\\
 &+&te^{i\kappa n}\left\{u_{\kappa,A}^{\beta}\delta_{\bar n,0}+u_{\kappa,B}^{\beta}\delta_{\bar n,1}\right\} \, ,
 \label{eq:right}
 \eer
 i.e., a superposition of two transmitted waves of $\alpha$ and $\beta$ polarization, where $t$ and $t'$ are the two transmission amplitudes. In the DW region, defined by $0 < n \leq 2 N$, the solution of Eq.~(\ref{schero_spin_wave}) with $\theta_n$ given by the DW solution is obtained numerically with boundary conditions imposed by the aforementioned ansatz evaluated at $n = 0$ and $n = 2 N +1$. The two-component character of the solution together with the two boundary conditions yields four linear equations from which the reflection and the transmission coefficients $r, r', t, t'$ can be obtained.

\section{Results and discussion}

In this section, we discuss our results of the scattering of spin waves of a DW and their analytical understanding based on the analytical structure of the transmission amplitudes.

\subsection{Scattering of spin waves with a smooth DW}
We set  $S_A=2.5$ and $S_B=2$ for the rest of this paper. The different values of spin on each sublattice cause the shape of the DW to deviate from the antiferromagnetic case, leading to corresponding changes in the scattering pattern. A schematic of a FiM DW and the corresponding transmission of RH and LH spin waves is shown in Fig.~\ref{sharp}. Let us first discuss the case of an atomically smooth DW depicted in Fig.~\ref{sharp}(a). An incoming RH (LH) spin wave with the energy of $\hbar\omega_{\alpha}$($\hbar\omega_{\beta}$) 
reverses its polarization upon transmission and appears as a LH (RH) spin wave on the right side of the DW. Since the energy of the incoming spin wave with a certain polarization is the same as the energy of the transmitted spin wave with the reversed polarization, the wave vector of the transmitted spin wave does not change (Figs.~\ref{sharp}(c),(d)). When $D/J\leq 0.1$, the transmission coefficients are almost the same for RH and LH spin waves. When $D/J\geq 0.2$, the transmission coefficient is significantly reduced for the RH spin wave in the entire range of $k$, but it remains relatively large for the LH spin wave in the range $0<k<0.3$, and the maxima of the transmission coefficient shifts towards lower wave vectors.

\subsection{Scattering of spin waves with a sharp DW}
Upon reaching the critical value of $D/J\approx 0.667$ [Eq.~(\ref{eq:dc})] and for all greater values of $D/J$,
the equilibrium configuration of the DW undergoes a transition to an abruptly sharp configuration as shown in Fig.~\ref{sharp}(b). In this case, the spin configuration is invariant under rotation around the $z$-axis. Owing to this symmetry, the state of polarization of the transmitted spin wave remains unchanged through the scattering process. As already mentioned, the spin wave dispersion for a given polarization is different on the left and right side of the DW; namely, on the left (right) side of the DW, the dispersion of a RH spin wave is given by $\omega_{\alpha}(\omega_{\beta})$. This leads to a change in the wave vector in order to maintain the frequency of the solution, i.e., $k \rightarrow \kappa$ with $\omega_\alpha(k) = \omega_\beta(\kappa)$, from which we obtain
 \ber 
 \kappa a =\cos ^{-1}\{\frac{1 }{4S_AS_BJ^2}(2JS_A+2DS_B-\hbar\omega_{\alpha})\nn\\ \times (2JS_B+2DS_A+\hbar\omega_{\alpha})  \}^{1/2}.
\eer
  By substituting  Eqs.~(\ref{eq:left}) and (\ref{eq:right}) into Eq.~(\ref{schero_spin_wave}) for an abrupt DW
 with two up-spins at the boundaries, the reflection amplitude
 of an incoming RH spin wave is

\begin{equation}
r^{(RH)}=-\frac{\mathcal{A}(k)}{\mathcal{A}(-k)} \, ,
\end{equation}
 where
 \ber
 \mathcal{A}(k) =-\frac{\hbar\omega_{\alpha}}{J}+2\frac{S_AD}{J}-\frac{2JS_AS_B e^{-ika}\cos ka}{\hbar\omega_{\alpha}+2S_A+2DS_B}\nn \\ +\frac{2JS_AS_B\cos \kappa a}{2(\hbar\omega_{\alpha}-2DS_B)\cos \kappa a + e^{i\kappa}(2S_A+2DS_B-\hbar\omega_{\alpha})} \, .\nn\\
 \label{eq:A(k)}
 \eer
The transmitted wave vector, as well as the transmission amplitude associated with an incoming LH spin wave can be obtained by changing $\omega_{\alpha}\rightarrow-\omega_{\beta}$ in the equations above.

Based on these analytical results, we plot the transmission coefficients against $k$ for different values of $D/J$. Let us first consider an RH spin wave as it gets scattered upon a sharp DW with two up-spin at the boundaries. As we can see from Fig.~\ref{sharp}(i), there is a gap in the transmission coefficient of the RH spin wave in which the wave experiences a total reflection due to a mismatch in the energy of the incoming and the transmitted spin waves. For larger values of $D/J$, the transmission coefficient and the gap decrease. In the case of an incoming LH spin wave, the transmission coefficient grows to be an order of magnitude larger than that of an RH spin wave. Furthermore, there is a cut-off in the transmission coefficient for high $k$, beyond which the spin wave is not able to pass through the DW.

\begin{figure}
	\centering
	\includegraphics[width=86mm]{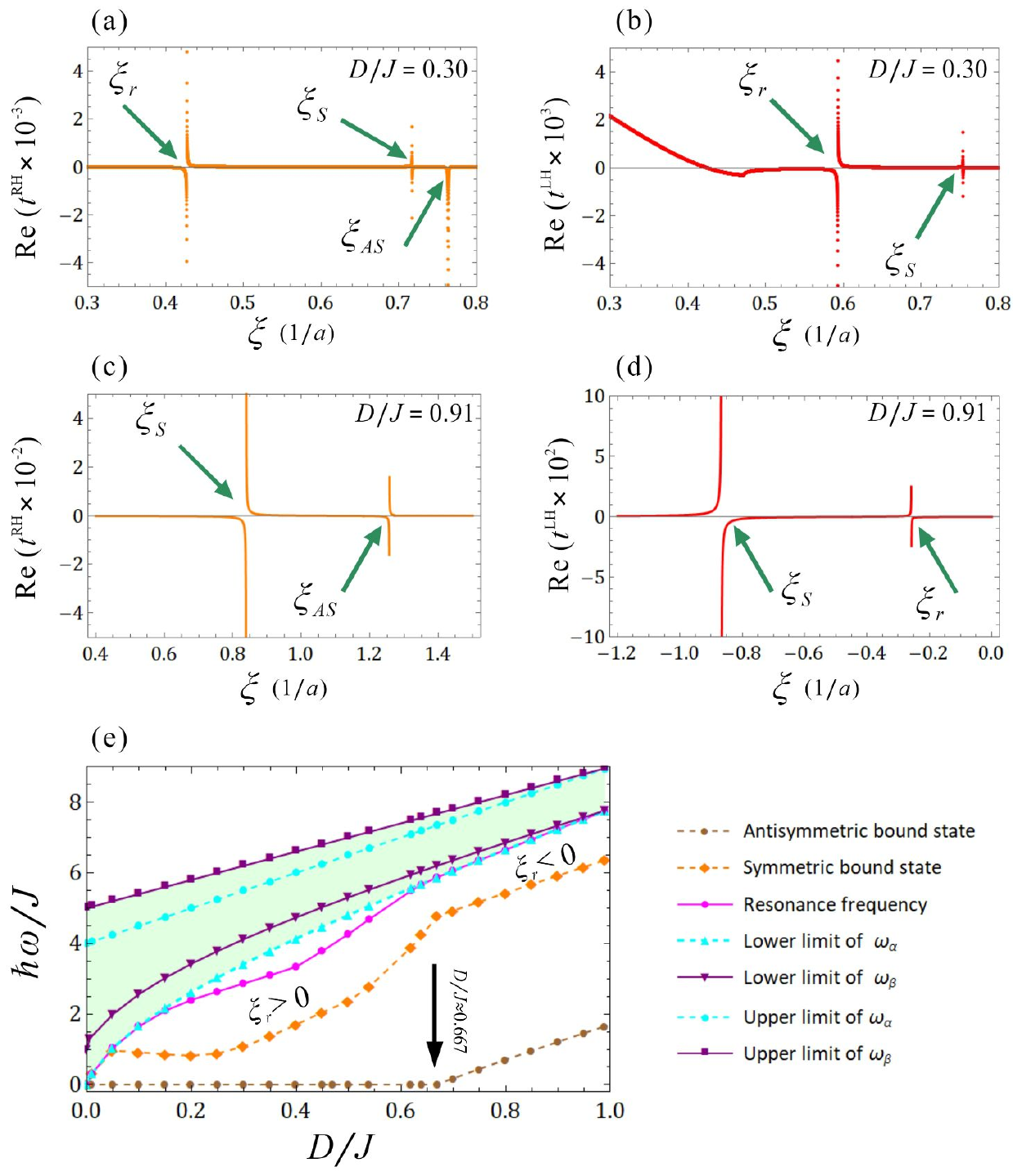}
	\caption{The real part of the transmission amplitude of (a) an RH spin wave and (b) an LH spin wave plotted as a function of wave vector $k=i\xi$ along the imaginary axis for $D/J=0.3$. (c),(d) The same as (a),(b) with $D/J=0.91$. (e) The spin-wave frequencies as a function of $D/J$. The shaded area represents the continuous spin-wave spectrum.}
\label{sharp_imaginary}
\end{figure}

\subsection{Analytical structure of the transmission amplitude}

We can qualitatively understand this scattering behavior of spin waves impinging upon a FiM DW by studying the analytic structure of the transmission amplitude for complex wave vectors along the imaginary axis, i.e., $k=i\xi$ where $\xi$ is real. The real part of the transmission amplitude is shown in Figs.~\ref{sharp_imaginary}(a-d). Let us first consider an incoming RH spin wave that impinges on a DW when $D/J$ is less than the critical value $d_c$. As we can see in Fig.~\ref{sharp_imaginary}(a), the transmission amplitude has three poles on the positive imaginary axis. The rightmost pole, $\xi_{AS}$, is the anti-symmetric bound state, which is associated with the Goldstone mode caused by axial symmetry of the Hamiltonian about the $z$-axis with $\omega_{AS}=0$. The middle pole, $\xi_{S}$, is associated with a symmetric oscillation of the spins around the equilibrium axis. The leftmost pole, $\xi_r$, is associated with an anti-symmetric oscillation of spins around their equilibrium and its corresponding frequency is located below the minimum of the continuous spin-wave spectrum.

Let us first consider the behavior of RH spin waves for $D/J<d_c$. The wave functions associated with the frequencies $\omega_r$, $\omega_S$ and $\omega_{AS}$ are bound states, in which the amplitudes of the oscillations decay exponentially away from the center of the DW.  According to Fig.~\ref{sharp_imaginary}(e) for a wide DW i.e., $D/J\ll 1$ the frequency of the symmetric bound state $\omega_S$ (orange curve) and the resonance frequency $\omega_r$ (pink curve) merge with the lower edge of the RH spin wave frequency spectrum $\omega_{\alpha}$, leading to a perfect transmission of the RH spin waves. As $D/J$ increases and the DW becomes narrower, $\omega_r$ (pink curve) begins to detach from $\omega_{\alpha}$ (blue curve), leading to a significant reflection of RH spin waves. For $D/J\gtrapprox 0.45$, the shape of the DW begins to change as the central spin begins to turn towards the easy axis. As a result, $\omega_r$ gets closer to $\omega_{\alpha}$ and finally at $D/J= d_c$ it equals $\omega_\alpha$ before detaching again at larger values of $D/J$. 

Let us now consider the behavior of LH spin waves, still for $D/J<d_c$.
The transmission amplitude of an incoming LH spin wave has only two poles on the positive imaginary axis as shown in Fig.~\ref{sharp_imaginary}(b): Notice that  $\xi_{AS}$  is absent. In addition, the lower limit of the LH spin wave frequency $\omega_{\beta}$ is far from the resonance frequency, and thus $\omega_r$ is not expected to influence the transmission coefficient.
   
Now let us turn to the large anisotropy case, $D/J>d_c$. The DW is atomically sharp and the polarization of the wave remains constant in the scattering process. In Figs.~\ref{sharp_imaginary}(c) and (d), we see that the transmission amplitude has only two poles on the imaginary $k$ axis: These two poles are on the positive semiaxis for RH waves (signifying bound states) but move to the negative semiaxis for LH waves (signifying anti-bound states).   
The resonance mode $\xi_r$, identified by its proximity to the lower limit of the spin wave continuum, is absent in the transmission amplitude of the RH spin wave. 




The difference in the transmission between the RH spin wave [Fig.~\ref{sharp}(i)] and the LH spin wave [Fig.~\ref{sharp}(j)] for a sharp DW can be understood as follows. For the antiferromagnetic case $S_A = S_B$, it can be shown analytically that the transmission probability for RH waves contains a factor $1/(\omega_{RH}+\omega_r)$, while the transmission probability for LH waves contains a factor $1/(\omega_{LH}-\omega_r)$ by examining Eq.~(\ref{eq:A(k)}) (see Ref.~\cite{faridi2022atomic} for the detailed discussions), which we expect to hold qualitatively also for FiMs. From these simple expressions, we see that, when the frequency of an LH spin wave is close to $\omega_r$ the transmission probability is on the order of the unitarity limit, i.e., 1 [see Fig.~\ref{sharp}(j)]. However, for an RH spin wave the denominator does not vanish and therefore the transmission probability remains well below the unitarity limit [see Fig.~\ref{sharp}(i)]. 

\begin{figure}
	\label{group}
	\centering
\includegraphics[width=0.6\columnwidth]{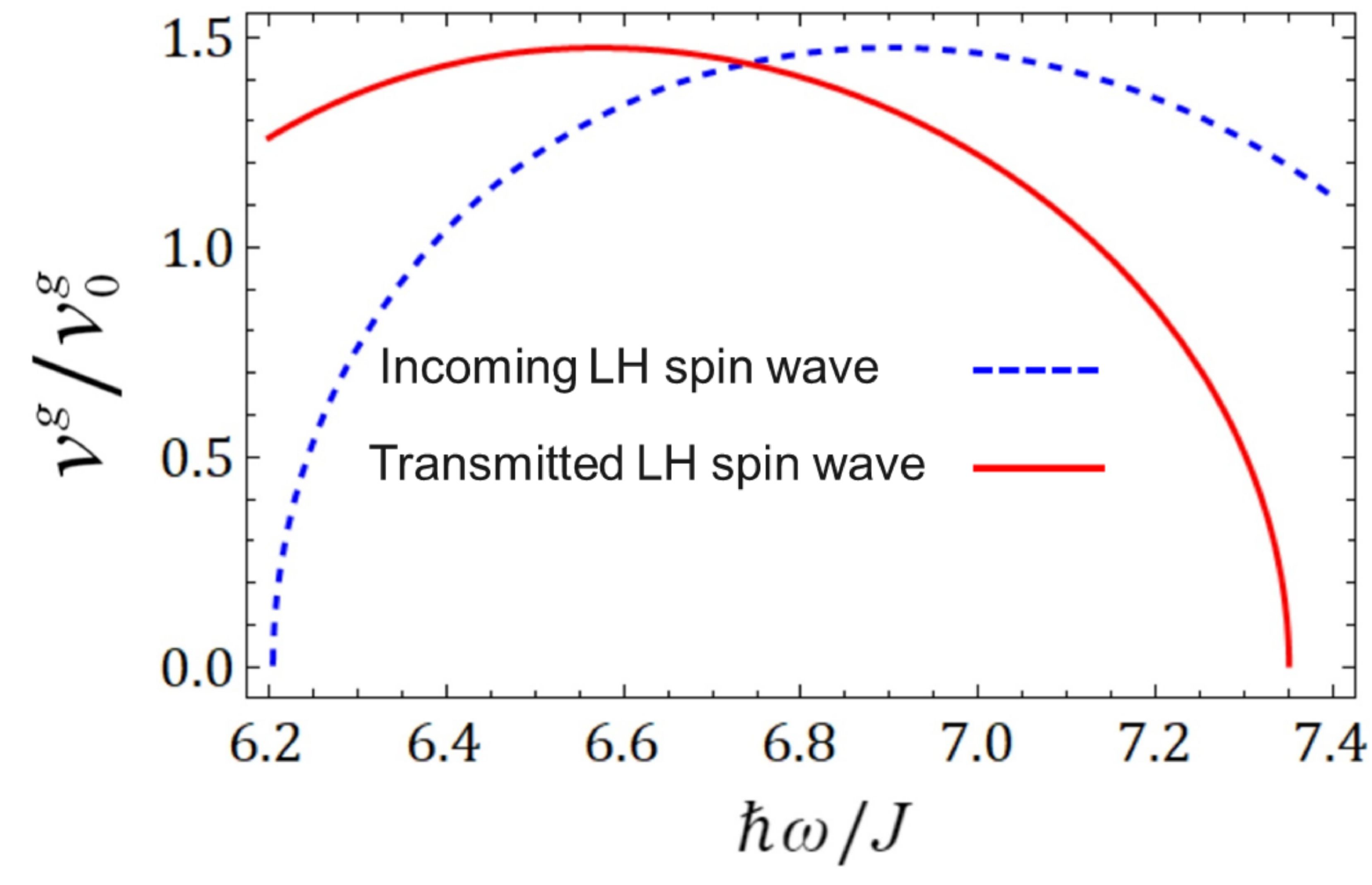}\caption{Group velocity of an LH spin wave coming from  the left side of the DW (blue dashed line) and group velocity of the transmitted LH spin wave on the right side of the DW (red line) as a function of the frequency.  Here, $D/J=0.67$ and $v_0^g = aJ/\hbar $.}
\label{fig:vg}
\end{figure}

\subsection{FiM DW as a spin-wave accelerator/decelerator} 

Our results show that an atomically sharp FiM DW could be used as a spin-wave accelerator and decelerator. According to Eqs.~(\ref{eq:omega-alpha}) and (\ref{eq:omega-beta}), and as illustrated in Figs.~\ref{sharp}(e) and (f), the dispersion relation changes between the left and the right sides of the DW for a given spin-wave polarization. For the spin-wave polarization that is transmitted through the DW, the frequency does not change, but the wave vector changes. For concrete and specific discussions, let us focus on the LH polarization. The wave vector $k$ increases after passing the DW, which leads to a change in the group velocity of the spin wave. The group velocity of an LH spin wave for $D/J=0.67$ is shown in Fig.~\ref{fig:vg}. The velocity of the transmitted spin wave (solid red curve) is increased relative to that of the incoming spin wave (dashed blue curve) between $6.20 \lesssim \hbar\omega/J \lesssim 6.75$, exhibiting the acceleration. For $\hbar \omega \gtrsim 6.75$, the spin wave is decelerated after passing the DW.

\section{\label{sec:level1}Conclusions}

In summary, we have investigated analytically and numerically the propagation and scattering of spin waves incident upon a FiM DW. We have shown that the specific form of a narrow  FiM DW leads to an enhanced filtration of the spin wave polarization relative to a homogeneous FiM spin chain. Furthermore, beyond a specific value of the anisotropy constant, the DW becomes abruptly sharp which preserves the state of the polarization of the spin wave but changes its wave vector upon transmission. The change in the wave vector gives rise to a change in the group velocity. This feature is exploited for us to propose an abruptly sharp FiM DW as a spin-wave accelerator or decelerator in spintronics devices. We envision generalizing our results for an atomically sharp DW in 1D FiMs to other magnetic solitons in 2D or 3D FiMs such as skyrmions and vortices may provide us with atomically small spin-wave controllers in higher-dimensional magnetic systems that can be realized neither with ferromagnets nor with antiferromagnets.

	\begin{acknowledgments}
	S.K.K. was supported by Brain Pool Plus Program through the National Research Foundation of Korea funded by the Ministry of Science and ICT (NRF-2020H1D3A2A03099291), by the National Research Foundation of Korea (NRF) grant funded by the Korea government (MSIT) (NRF-2021R1C1C1006273), and by the National Research Foundation of Korea funded by the Korea Government via the SRC Center for Quantum Coherence in Condensed Matter (NRF-RS-2023-00207732).
	\end{acknowledgments}

\appendix \section{The critical value of $D/J$}

In this appendix, we find the critical value of $D/J$ at which a FiM DW becomes abruptly sharp by using the fact that above the critical point, the zero frequency Goldstone mode cannot exist.  To do so, we can consider an evanescent spin wave on top
of an abruptly sharp domain wall by switching to the imaginary wave vector. By employing the
equation of motion and the structure of an evanescent wave on top of a 
DW, we obtain the following equations for allowed eigenfrequencies of spin
wave including symmetric and antisymmetric bound states as well as the resonance frequency.

First we introduce the Ansätze for  $n\leq0$:
\begin{eqnarray} \label{e-wave}
	\psi_n= r\left\{u^{(\alpha)}_{k,A}\delta_{\bar n,0}+u^{(\alpha)}_{k,B}\delta_{\bar n,1}\right\} e^{k na} \nonumber\\
	+r'\left\{u^{(\beta)}_{\kappa,A}\delta_{\bar n,0}+u^{(\beta)}_{\kappa,B}\delta_{\bar n,1}\right\} e^{\kappa na},
 \label{leftside}
\end{eqnarray}
and for $ n\geq1$
\begin{eqnarray}
	\psi_n=t'\left\{u^{(\alpha)}_{k,A}\delta_{\bar n,0}+u^{(\alpha)}_{k,B}\delta_{\bar n,1}\right\} e^{-k na}\nonumber \\+t\left\{u^{(\beta)}_{\kappa,A}\delta_{\bar n,0}+u^{(\beta)}_{\kappa,B}\delta_{\bar n,1}\right\} e^{-\kappa na}\,. 
 \label{rightside}
\end{eqnarray}

By substituting Eqs.~(\ref{leftside}) and (\ref{rightside}) in the equation of motion at the boundaries $n=0$ and $n=1$:

\ber \label{EOMpsi0}
	H_{0\alpha,0\beta} \psi_0+H_{0\alpha,-1\beta} \psi_{-1}	+H_{0\alpha,1\beta} \psi_1&=&\hbar\omega \psi_0, ~~n=0 \, , \nn\\
	H_{1\alpha,1\beta} \psi_1+H_{1\alpha,0\beta} \psi_{0}	+H_{0\alpha,2\beta} \psi_2&=&\hbar\omega \psi_1, ~~n=1 \, , \nn\\
	 \eer
the following equation is obtained:
\begin{widetext}
\ber
\left(\frac{2 S_A S_B e^{-ka } \cosh ka }{2 d S_B+2 S_A+\hbar\omega }-2 d S_A+\hbar\omega \right) \left(\frac{2 S_A S_B e^{-\kappa a } \cosh \kappa a }{2 d S_A+2S_B+\hbar\omega }-2 d S_B+\hbar\omega \right)-S_A S_B=0 \, .
\eer
\end{widetext}
Setting $\omega=0$ and after  simplification, we end up with a third degree algebraic equation of $d$: 
\begin{equation}
	-3 S_AS_B d^3-2(S_A^2+S_B^2)d^2+S_AS_B d+S_A^2+S_B^2=0 \, .
\end{equation} 
The analytical solution to this equation is given by:
\ber
d_c = \sqrt[3]{\Delta_0 + \sqrt{\Delta_0^2 + \Delta_1^3}}+ \sqrt[3]{\Delta_0 - \sqrt{\Delta_0^2 + \Delta_1^3}} -\frac{2}{9}\mathcal{S} \, ,
\label{eq:dc}\nn\\
.
\eer
where $\mathcal{S}=(S_A/S_B+S_B/S_A$), $\Delta_0=-8\mathcal{S}^3/27^2+7\mathcal{S}/54$, $\Delta_1=-1/9-4\mathcal{S}^2/81$.

	\nocite{*}
	
	\bibliography{./fimreferences}
	
\end{document}